\documentstyle[pre,aps]{revtex}
\begin{document}
\draft
\title{Non equilibrium electronic distribution in single electron devices}
\author{N. Garc{\'\i}a$^{\dag}$ and F. Guinea$^{\ddag}$}
\address{
$^{\dag}$
Laboratorio de F{\'\i}sica de Sistemas Peque\~nos.
Consejo Superior de Investigaciones Cient{\'\i}ficas.
Serrano 144. 28006 Madrid. Spain. \\
$^{\ddag}$ 
Instituto de Ciencia de Materiales.
Consejo Superior de Investigaciones Cient{\'\i}ficas.
Cantoblanco. 28049 Madrid. Spain.  \\
}
\date{\today}
\maketitle 
 
\begin{abstract}
The electronic distribution in devices with sufficiently small
dimensions
may not be in thermal equilibrium with their surroundings. 
Systems where the occupancies of electronic states are solely determined by
tunneling processes are analyzed.  It is shown that the effective
temperature of the device may be higher, or lower, than that of its
environment, depending on the applied voltage and the energy dependence
of the tunneling rates. The I-V characteristics become asymmetric. 
Comparison with recent experiments is made.
\end{abstract}


%\narrowtext
In small devices, the coupling between electrons and the lattice is
suppressed.
At sufficiently low
temperatures, only long wavelength accoustical phonons are excited.
If the electrons are localized in a region much smaller than the
wavelength of the phonons, the interaction between the electrons and the
phonons is significantly reduced\cite{heating}.
As a result, the electron temperature may be different from that of the
surrounding medium.

Usually, the electron temperature in small devices tend to be higher
than that of the environment,
because of dissipation at the device\cite{heating}
(see also\cite{heating2}), via shake-up
processes. This effect can
explain a number of experiments\cite{exph,exph2}.

In the following, we analyze the electronic distribution in a
small device, when it is controlled by the electronic exchange with
the external leads. The general equations which describe the 
single level occupancies were first discussed in\cite{AKL,CWJB}.
These authors considered mostly the inluence of a non negligible single
level spacing in the I-V characteristics of semiconductor nanostructures.  
In these systems, the effects of the energy dependence of the transmission
through the barrier, or of the density of states in the external leads
is not important. We will  generalize the previous work to situations
where energy dependent tunneling rates also contribute to modify the
electronic distribution. As it will be seen below, an  
energy dependent tunneling rate, $T ( \epsilon )$, can modify significantly
the electronic distribution in the central electrode of
a single electron device, if $T '  ( \epsilon ) / T ( \epsilon )$ is comparable
to the inverse of the temperature  at which the device is operated.
Such a situation can be realized when an electrode has a strongly dependent
density of states, or if the tunneling process takes place close to the top
of a barrier.  The first case is relevant to the experiments reported 
in\cite{expc}, where one of the electrodes is made of graphite, and to
the experiments presented in\cite{expsc}, where the electrodes are made of 
superconducting Al, and tunneling processes take place near the gap edges.
Asymmetric I-V characteristics have also been reported in\cite{STM}. Some
features of these experiments are also consistent with the work reported here.

A particularly interesting case is presented in\cite{expc}. The observed
Coulomb staircase can only 
be fitted by the orthodox theory\cite{KS,Likharev,stairs},
if an effective temperature of $\sim 170$K is assumed, although the
experiment is performed at room temperature. 
A large temperature difference between the central electrode in the device 
used in\cite{expc} and its surroundings cannot be ruled out.
The central island is isolated by a Langmuir-Blodgett film from the
rest of the system. A simple estimate of the temperature difference can be
made by assuming that the energy dissipated in the circuit has to be
carried away through the film. Then:

\begin{equation}
\frac{\partial E}{\partial t} = I V \sim \kappa L ( T_{env} - T_{elec} )
\label{estimate}
\end{equation}
where $\kappa$ is the thermal conductivity of the film, $L$ is the linear
size of the electrode, $T_{env}$ is the temperature of the environment,
and $T_{elec}$ is the temperature of the electrode.
$I$ and $V$ are the intensity and the voltage acting on the device.
Typical values of $\kappa$ for organic films, such as Mylar, are 
$\sim 10^{-4}$ W / ( m K ). The size of the electrode is
$L \sim$ 20\AA. Using the values for $I$ and $V$ reported in\cite{expc},
we find that $T_{env} - T_{elec} \sim 10^2$K. Note that the temperature
gradient supported by a good thermal insulator, such as an organic film, 
is much larger than the one expected in a system 
where heat can be dissipated  through metallic regions ( $\kappa \sim 1$
W / ( m K )).

A different way to manipulate the electronic distribution in a
mesoscopic device has been discussed in\cite{niu}. Our  analysis differs
from that in\cite{niu} because we will study systems in which the 
electronic distribution is not in thermal equilibrium.

The orthodox theory of Coulomb blockade\cite{KS,Likharev,stairs}
analyzes the transport processes in small capacitance devices by means
of rate equations. 
These equations assume that the different elements of
the device are in thermal equilibrium. The rate equations allow us to
obtain the (time dependent) number of electrons in each individual
electrode.
We  will relax the hypothesis of electronic thermal
equilibrium, as in\cite{CWJB,AKL}. We assume that transport 
processes determine, not only the
evolution of the charge, but the fluctuations in the occupancies of
individual levels as well.  The influence of other
effects,  such as electron-electron collisions or the coupling to the
lattice, will be discussed later.

We study a small island, coupled by two junctions to leads connected to
a battery which maintain a voltage difference, $V$. Each junction is
characterized by a capacitance and a resistance, $C_{1,2}$ and
$R_{1,2}$.
The electronic temperature in the leads, $T$,  is fixed by the cryostat.
We will assume that the charge in the island fluctuates between two
states, with $N$ and $N + 1$ electrons. 
The island has an effective charging energy of $e^2 / (
C_1 + C_2 )$\cite{stairs}. When the island has charge $N$, the voltages
at each junctions are\cite{stairs}:
\begin{eqnarray}
V_1 &= &\frac{e N}{C_1 + C_2} + V \frac{C_2}{C_1 + C_2} \nonumber \\
V_2 &= &- \frac{e N}{C_1 + C_2} + V \frac{C_1}{C_1 + C_2}  
\label{voltage}
\end{eqnarray}
These voltages determine the tunneling rate of electrons to and from the
leads.

We label the probability that the island is in a state with $N$
electrons in the levels $i_1 , i_2 , ... , i_N$ as
$n_{i_1 , i_2 , ... , i_N}$, and we use a similar convention for the 
cases with $N + 1$ electrons. The probability for an electron from lead
1 to jump into  a state of energy $\epsilon$ in the island 
is proportional to the occupancy of a given state at that 
energy,  $n ( V_1 + \epsilon )$,
times the transition rate across junction 1, $t_1 ( V_1 + \epsilon )$,
where $n ( \epsilon )$ is the Fermi-Dirac distribution at temperature $T$.
For simplicity, we include density of states effects in the definition
of the $t_i$'s.
Analogously, the probability that an electron of energy $\epsilon$ leaves
the island through junction 1 is proportional to  $1 - n ( V_1 +
\epsilon )$ times $t_1 ( V_1 + \epsilon )$. We define, in this way, the
quantities:
\begin{eqnarray}
p_1^{in} ( \epsilon ) &= &t_1 ( V_1 + \epsilon )  n ( V_1 + \epsilon )
\nonumber \\
p_1^{out} ( \epsilon ) &= &t_1 ( V_1 + \epsilon ) [ 1 - n ( V_1 +
\epsilon ) ] \nonumber \\
p_2^{in} ( \epsilon ) &= &t_2 ( V_2 + \epsilon )  n ( V_2 + \epsilon )
\nonumber \\
p_2^{out} ( \epsilon ) &= &t_2 ( V_2 + \epsilon ) [ 1 - n ( V_2 +
\epsilon )  ] 
\label{probabilities}
\end{eqnarray}
Then, the probabilities of finding a given
electronic configuration in the island obey:
\begin{eqnarray}
\frac{\partial n_{i_1 , i_2 , ... , i_N}}{\partial t} &=
&\sum_{i_k \ne i_1 ... i_N} [ p_1^{out} (  \epsilon_{i_k} ) + 
p_2^{out} ( \epsilon_{i_k} ) ] n_{i_1 , ... , i_N , i_{N + 1} = i_k}
\nonumber \\ &-
&n_{i_1 , i_2 , ... , i_N} \sum_{i_k \ne i_1 , ... , i_N}
[ p_1^{in} ( \epsilon_{i_k} ) + p_2^{in} ( \epsilon_{i_k} ) ] \nonumber
\\
\frac{\partial n_{i_1 , i_2 , ... , i_{N + 1}}}{\partial t} &=
&\sum_{i_k = i_1 ... i_{N + 1}} [ p_1^{in} (  \epsilon_{i_k} ) + 
p_2^{in} ( \epsilon_{i_k} ) ] n_{i_1 \ne i_k , ... , i_N \ne i_k } 
\nonumber \\ &-
&n_{i_1 , i_2 , ... , i_{N + 1}} \sum_{i_k = i_1 , ... , i_{N + 1}}
[ p_1^{out} ( \epsilon_{i_k} ) + p_2^{out} ( \epsilon_{i_k} ) ] 
\label{rate}
\end{eqnarray}
These equations are a straightforward generalization to the ones
formulated in\cite{CWJB,AKL} to the case of energy dependent 
tunneling rates. They
admit the stationary solution:
\begin{equation}
n_{i_1 , i_2 , ...} = C \prod_{i_k = i_1 , ...
} \frac{p_1^{in} ( \epsilon_{i_k} ) +
p_2^{in} ( \epsilon_{i_k} )}{p_1^{out} ( \epsilon_{i_k} ) + p_2^{out} (
\epsilon_{i_k} )} = C \prod_{i_k = i_1 , ...}  f ( \epsilon_{i_k} )
\label{occupancies}
\end{equation}
where $C$ is a normalization constant and:
\begin{equation}
f ( \epsilon ) = \frac{p_1^{in} ( \epsilon ) + p_2^{in} ( \epsilon
)}{p_1^{out} ( \epsilon ) + p_2^{out} ( \epsilon )}
\label{factor}
\end{equation}
Charging effects are built in
into eq.(\ref{occupancies}) through the dependence of the $p_i$'s on
$V_1$ and $V_2$.

Each factor in the product in eq.(\ref{occupancies}) can be interpreted
as a Boltzmann weight which determines the occupancy of the
corresponding level. The previous scheme can be extended to situations
where more than two charge states are involved. Then, each pair of
charge states, $N$ and $N + 1$, determine a set of equations of the form
(\ref{rate})\cite{stairs}. The quantities $p_i^{in}$ and 
$p_i^{out}$ depend on $N$ and $N + 1$. A simple solution, like
(\ref{occupancies}), is no longer possible. A closed
expression can be obtained in terms of sums over all charging
processes which lead to a charge state $N$, with occupancies $\{ i_k \}$,
from the minimum charge state included in the model, $N_{min}$.
Finally, in the absence of charging processes, the voltages $V_1$ and
$V_2$ are independent of the charge state of the central islands. 
The solution of the rate equations (\ref{rate})  is given by 
(\ref{occupancies}). The calculation of $V_1$ and $V_2$, however,
becomes more complicated, as they are determined by the junction
resistances, and a self consistency loop is required.

In the following, we will describe the electronic distribution by means
of $f ( \epsilon )$, (\ref{factor}).  This scheme  is valid for
voltages in the flat portions of a
Coulomb staircase, when only two charge states
are involved. Close to the steps, the island fluctuates between three
charge states, $N , N + 1$ and $N + 2$. The chemical potential for state
$N + 2$ is slightly above the chemical potential of one of the leads.
Then, the 
rates $p^{in}$ and $p^{out}$ for processes which take the island in or
out of state $N + 2$  depend more weakly on energy than the processes which
relate states $N$ and $N + 1$. In this case, a description of the
occupancies in the island by a function $f ( \epsilon )$ which depends
only on the rates between $N$ and $N + 1$ is adequate.

The distribution described by $f ( \epsilon )$ is not in thermal
equilibrium. This is reflected in the fact that the quantity
defined as $\beta ( \epsilon ) = - \partial \log  [ p ( \epsilon ) ]
/ \partial \epsilon$ is not independent of $\epsilon$, although the variations
of $\beta ( \epsilon )$ will be less pronounced than those of $p ( 
\epsilon )$. If $\beta ' ( \epsilon ) \ll \beta^2 ( \epsilon )$,
an approximate thermal distribution can be
defined in an energy range $\sim \beta^{-1}$ around $\epsilon$.
The effective temperature is $\beta^{-1}$. 
In the central electrode, unless $\beta$ has a very unusual dependence on
$\epsilon$, te lowest lying states will be fully occuppied, and the
highly excited ones, empty. Hence, if $\beta ' \ll \beta^2$
around the equilibrium Fermi level, the new distribution is
completely specified by $1 / ( k_B T_{eff} ) = \beta ( \epsilon_F )$.
$T_{eff}$, defined in this way, gives the best fit to the
level occupancies within the island by a thermal distribution. In
principle, it depends on $\epsilon$. 

The following properties of $T_{eff}$ are easy to prove:

(i) $T_{eff} = T$ if $V = 0$, for any $t_{1,2} ( \epsilon )$.

(ii) $T_{eff} \rightarrow T$ if $t_1$ and $t_2$ are independent of 
energy, and $t_2 \ll t_1$, or $t_1 \ll t_2$. The thermalization
of the central electrode in a very asymmetric junction was discussed
in\cite{AK}. 

(iii) $T_{eff} = T$, if $t_1 = t_2$. The lack o corrections
to the equilibrium distribution in a symmetric junction, to lowest
order in the applied voltage, was discussed in\cite{CWJB}.

(iv) If $V \ne 0$ and $t_1$ , $t_2$ are independent of energy,
$T < T_{eff}$. $T_{eff}$ reaches a maximum for $V_1 <
\epsilon < V_2$. If $T_{eff} \rightarrow \infty$,
all electronic
configurations with a given $N$ are equally likely. This situation may
have been observed experimentally\cite{teff}. Note, however, that, if
$T_{eff} \rightarrow \infty$, $\beta ' \gg \beta^2$. 
The electronic distribution cannot be well described by
$T_{eff}$ in an energy range comparable to $T_{eff}$ itself, in this case.

We now consider the situation when the $t_i$'s depend on $\epsilon$. 
We study the case $t_2 \ll t_1$, which is a
neccessary condition for the observation of a Coulomb
staircase\cite{stairs}. Then, the leading effects associated to
$t_1 ' , t_2 '$ give:
\begin{eqnarray}
\frac{1}{k_B T_{eff}} &\approx  &\frac{1}{k_B T} - \left( \frac{t_1 '}{t_1} -
\frac{t_2 '}{t_2} \right) \frac{t_2}{t_1} \left[ \frac{n ( V_2 +
\epsilon )}{n ( V_1 + \epsilon )} 
- \frac{1 - n ( V_2 + \epsilon )}{1 - n ( V_1 + \epsilon )}\right] , 
\nonumber \\
& &\, \, \, \, \, \, \, \, \, \, \, \, \, \, \,
\frac{t_2}{t_1} \ll {\rm min} [ n ( V_1 + \epsilon ) , 1 -
n ( V_1 + \epsilon ) ] \nonumber \\
\frac{1}{k_B T_{eff}} &\approx  &\frac{t_1 '}{t_1} ,
\, \, \, \, \, \, \, \, n ( V_1 + \epsilon ) \ll \frac{t_2}{t_1} \nonumber \\
\frac{1}{k_B T_{eff}} &\approx  &-  
\frac{t_1 '}{t_1} , 
\, \, \, \, \, \, \, \, 1 - n ( V_1 + \epsilon ) \ll \frac{t_2}{t_1} 
\label{cooling}
\end{eqnarray}
The most favorable case for cooling occurs when $n ( \epsilon + V_1 )
\ll t_2 / t_1$. Then, $T_{eff} \sim {\rm min} [ T , t_1 ' / ( k_B t_1 )
]$.
If the junctions are tunnel barriers, 
$t_i ( \epsilon_i ) \propto e^{- C \frac
{\sqrt{m ( V_0 - \epsilon_i)} l}{\hbar}}$,
where $C$ is a constant of order unity,
$V_0$ is the height of the barrier, $m$ is the mass of the
electron and $l$ is the length of the barrier. Then,
\begin{equation}
\frac{t_i '}{t_i} \sim \sqrt{\frac{m l^2}{\hbar^2 ( V_0 - \epsilon_i )}}
\label{barrierest}
\end{equation}
Typical values for $m , V_0$ and $l$ give $t_i ' / t_i \sim 1$
(eV)$^{-1}$.  Density of states effects, in a normal metal, give
modulations to the $t_i$'s of similar magnitude.
Hence, the $\epsilon$ dependence of the tunneling rates 
should not induce large effects in conventional SET's,
unless the Fermi surface of the central island lies close to the top of
one of the barriers.

Significant changes in the $T_{eff}$'s are expected if the tunneling
rates, or the density of states in the substrate, depend strongly on
energy. This is the case in the experiments reported in\cite{expsc}.
The Fermi energy of the central island is aligned with the edge of the
gap of one of the superconducting electrodes. Thus, the process studied
here can contribute to the cooling observed in\cite{expsc}.
A different situation is the setup used 
in \cite{expc}. One of the substrates is
graphite, which is a semimetal.
The density of states per atom,  
near the Fermi level, increases roughly linearly
with energy, $D ( \epsilon )  \sim | \epsilon - \Delta | / \epsilon_0^2$,
where $\epsilon_0 \sim 4$eV, and $\Delta \sim
10^{-4}$eV\cite{graphite}. Hence, $k_B T_{eff} \sim  D / D' \sim
| V_1 - \Delta |^{-1}$. Near the steps of the Coulomb staircase, the
Fermi level of the island is close  to that of the substrate, so that
$V_1$ measures the distance to the step edge. The effective temperature
can be well below that of the cryostat.
Note that, for a tunneling rate which depends linearly on energy,
$\beta ' ( \epsilon ) \sim \beta^2 ( \epsilon )$, so that the 
characterization of the distribution by a single parameter, $T_{eff}$,
is only an approximation. The overall effect on the Coulomb staircase,
however, is a significant sharpening of the steps, in agreement 
with\cite{expc}.

\begin{figure}
\caption{(a) Combination of external voltage and barriers
for electron cooling (high energy electrons leave the central island
faster than the low energy ones). (b) Combination of external voltage
and barriers for electron heating (high energy electrons arrive to the
island faster than the low energy ones).}
\label{dvcool}
\end{figure}
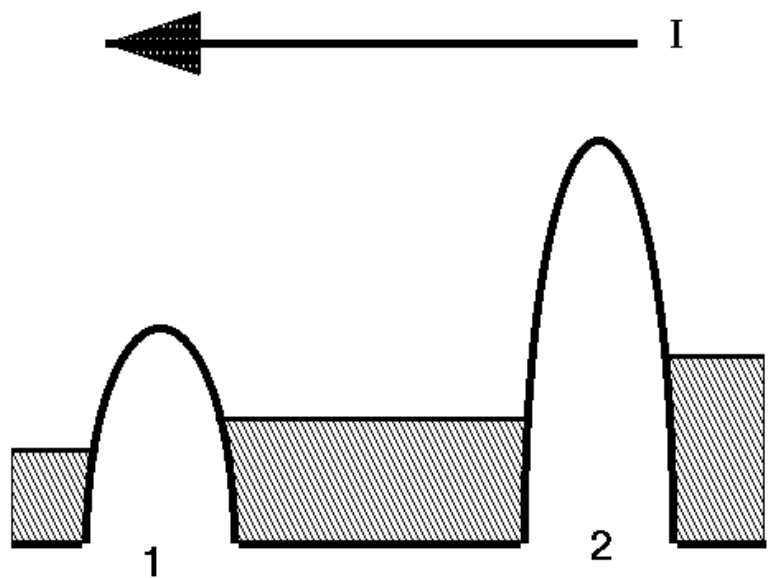
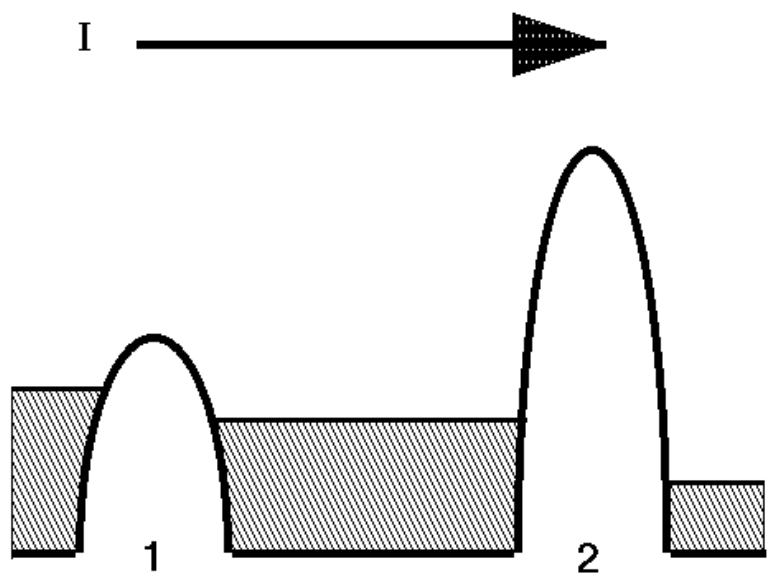

The general situation of energy dependent barriers, or density of
states, is schematically shown in fig.(\ref{dvcool}). 
Intuitively, if the
value of $t_1 ( \epsilon )$ rises sufficiently fast with $\epsilon$,
and electrons flow into lead 1,
the high energy electrons are blown away from the island. 
If the applied voltage is reversed, hot electrons are pushed
into the island, where they are blocked by the high resistance
junction, 2. The I-V characteristics will be highly asymmetrical.
A population inversion, in which the high energy
states show larger occupancies than the low energy ones is
also possible. This situation corresponds to $T_{eff} < 0$. 

So far, we have assumed that the electron distribution within the island
is entirely determined by the tunneling processes to and from the leads.
At the beginning of the article we presented an estimate of the  efects o
heat conduction across the surrounding medium on the temperature of
the central electrode. We need also to consider electron-electron
interactions, among the electrons in the central electrode and those
in the leads, and within the electrode itself.
In the orthodox theory of Coulomb blockade, tunneling
processes are assumed to involve no dissipation. The relaxation of this
approximation leads to shake up effects, and heating\cite{heating}.
These processes impose a lower bound on the island electronic
temperature, $T_{su}$, which is a fraction of the Coulomb energy, $E_c$. Taking
$k_B T_{su} \sim E_c / 8$\cite{heating} and $E_c \approx
0.12$eV\cite{expc}, we find $T_{su} \sim 170$K. 

Finally, electron-electron interactions within the island will tend to
thermalize the electronic distribution. The thermal 
energy in the island is
proportional to $T_{eff}$, taken 
at the Fermi level. Hence, the equilibrium
temperature will not be very different from this value.

One of us (N. G.) is thankful to  M. Jonson and R. I. Shekter for
helpful discussions.

\end{document}